\documentclass[10pt]{iopart}
 \usepackage{hyperref}
 \usepackage{color}
 \usepackage[utf8]{inputenc}
 \usepackage{bm}
 \usepackage{graphicx, subcaption}
 \usepackage{latexsym}
 \usepackage{pifont}
 \usepackage{amssymb}
 \usepackage{hyperref}
 \usepackage{titlesec}
 \expandafter\let\csname equation*\endcsname\relax
 \expandafter\let\csname endequation*\endcsname\relax
 \usepackage{amsmath}
 \usepackage{physics}
 \usepackage{listings}
 \usepackage{verbatim}
 \usepackage{geometry}
 \usepackage{booktabs}
 \usepackage{enumitem}
 \usepackage[normalem]{ulem} 
 \usepackage{breqn}
 \usepackage[toc,page]{appendix}
 
 \definecolor{red}{rgb}{1.0,0.0,0.0}
 \definecolor{gre}{rgb}{0.0,1.0,0.0}
 \definecolor{blu}{rgb}{0.0,0.0,1.0}

 \definecolor{ora}{rgb}{1.0,0.5,0.0}
 \definecolor{gra}{rgb}{0.0,0.5,1.0}

 \newcommand{\change}[2]{\sout{#1}\red{#2}}
 
 \newcommand{\ba}{\begin{abstract}}
 \newcommand{\ea}{\end{abstract}}
 \newcommand{\be}{\begin{equation}}
 \newcommand{\ee}{\end{equation}}

 \newcommand{\eg}{\textit{e}.\textit{g}. }

 \newcommand{\rom}[1]{\uppercase\expandafter{\romannumeral #1 \relax}}

 \newcommand{\Eqn}[1]{Eq.(\ref{eq:#1})}

 \newcommand{\ds}{\displaystyle}
 
 \newcommand{\vect}[1]{\mathbf{#1}}
 
\usepackage{esint}
\renewcommand{\change}[2]{#2}

\hypersetup{
     colorlinks   = true,
     linkcolor    = blue,
     citecolor    = blue,
     urlcolor     = blue
}
\begin{document}
\title{Stellarator coil design using cubic splines for improved access on the outboard side}
\author{Nicola Lonigro$^1$, Caoxiang Zhu$^2$}
\address{$^1$ Department of Physics and Astronomy, University of Padua, Via Francesco Marzolo 8, Padua (PD), Italy}
\address{$^2$ Princeton Plasma Physics Laboratory, Princeton University, P.O. Box 451, New Jersey 08543, USA}
\ead{\mailto{nicola.lonigro@studenti.unipd.it, czhu@pppl.gov}}

\begin{abstract}
In recent years many efforts have been undertaken to simplify coil designs for stellarators due to the difficulties in fabricating non-planar coils. The FOCUS code removes the need for a winding surface and represents the coils as arbitrary curves in 3D. In the following work, the implementation of a spline representation for the coils in FOCUS is described, along with the implementation of a new engineering constraint to design coils with a straighter outer section. The new capabilities of the code are shown as an example on HSX, NCSX, and a prototype quasi-axisymmetric reactor-sized stellarator. The flexibility granted by splines along with the new constraint will allow for stellarator coil designs with improved accessibility and simplified maintenance.   
\end{abstract}


\section{Introduction}
Stellarators are 3D plasma configurations where both the poloidal and toroidal field is generated by the magnetic coils thus avoiding the instabilities driven by large plasma currents \cite{Fri}. To generate the required magnetic field, the coils need to twist around the plasma and this leads to complex shapes which are difficult to manufacture. The typical approach for designing an optimized stellarator consists of determining a magnetic configuration with good confinement properties, which can be achieved using the STELLOPT suit of codes \cite{osti_1617636}, and then design a coil set able to reproduce the desired magnetic configuration. 

Stellarator Coil design was initially performed in NESCOIL\cite{Merkel_1987} by solving for the current potential on a winding surface which was then discretized in a finite number of filaments. Later, this approach was improved by the codes NESVD \cite{Pomphrey_2001} and  REGCOIL\cite{Landreman_2017}. The need to directly control the shape of the coils lead to the creation of optimization codes such as ONSET \cite{ONSET}, COILPLOT \cite{coilopt}, and COILPLOT++ \cite{Gates_2017}. In this type of code, the coils are represented as filaments on a toroidal winding surface and a non-linear optimization is performed on this surface to obtain the final shape for the set of coils. A drawback of this kind of approach is that the best winding surface to use for a specific problem is not known a priori and the use of an unsuitable winding surface will lead to suboptimal coils.
FOCUS \cite{zhu01} is a numerical code able to design coils without the need for a winding surface but instead describing the coils directly as arbitrary curves in 3D, making it able to sample a wider range of possible configurations. There are also other recent advances in stellarator coil design to address specific problems \cite{Lobsien2018, Yamaguchi2019, Giuliani2020}.

Large access on the outboard side is vital for remote maintenance in future fusion reactors.
This is even more critical for stellarators as non-planar coils usually have limited space.
COILOPT++ showed the improvements obtained by using a spline representation for the coils.
It has the ability to implement local constraints in real space in a straightforward way.
For example, it is possible to fix the outer part of the coils to be circular arcs to create space for ports and allow easier access to the modules of the blanket\cite{Coilplot}.
In previous versions of FOCUS the coils were described using a Fourier representation, which is a global representation and so less suited to implement constraints on just a portion of the coils. Also, the Fourier representation is not efficient in describing straight sections.
In this paper, we add a spline representation for the coils, along with a new engineering constraint to design coils with straighter outer sections.
The differences between this work and the previous COILOPT++ approach are as follows.
COILOPT++ represents coils as planar curves on a given winding surface, while we follow the logic of FOCUS and adopt a cubic B-spline representation.
The coils can be freely moved in 3D space and thus we could explore more possible designs.
In addition, we use analytical derivatives during the coil optimization to improve both the speed and accuracy.
With this improvement, FOCUS is now able to design coils with straighter outboard parts which is extremely favorable for stellarator rectors.

This paper is organized as follows. In section \ref{sec:spline} a brief overview of spline representations is given and in section \ref{sec:opt} the optimization procedure is described. The improved coils for the HSX and NCSX stellarators along with a quasi-axisymmetric reactor-like configuration obtained using the new additions to the code are described in section \ref{sec:applications}. The conclusions and further improvements to the code are discussed in section \ref{sec:concl}. 

\section{Cubic B-spline representation}\label{sec:spline}
When describing a curve different parametrizations can be used with some being more appropriate for a certain problem than others. A parametrization is a set of functions that describe the coordinates of the curve in real space as a function of a parameter $x$. In a polynomial representation the coordinates of the curve are a polynomial function of the parameter. While an arbitrarily complex curve can be described using an high enough order for the polynomial, the resulting parametrization can be unstable and small changes in the polynomial coefficients can lead to large variations in the represented curve. 
A spline representation avoids this problem by stitching together many low-order polynomials in series. This allows representing an arbitrary complex curve by increasing the number of polynomial sections joined together instead of the degree of each polynomial. The points where the polynomials meet in the parameter space are called \textit{knot points} and the interval between two knot points is called \textit{knot interval}. 
Many possible spline representations exist\cite{splines} and the most appropriate one depends on the specific problem at hand. Using a spline of order 3, only two of the following properties can be achieved: $\mathbb{C}^2$ continuity, interpolation of all the control points or local control.
For this work, a local representation is preferred as it allows to modify one section of the curve while leaving the rest of the coil unchanged. 
This leads to discarding  Natural, Hermite, Catmull-Rom and Cardinal cubic splines due to their lack of locality while B\'ezier curves were discarded due to their lack of $\mathbb{C}^2$ continuity at the connection points between neighboring sections. B-splines were selected due to both their locality and $\mathbb{C}^2$ continuity even though the curves do not pass through the control points used to define them, slightly complicating their interpretation.
In the case of B-splines, the coefficients used in the definition take the name of \textit{control points} $C_i$ and by changing their coordinates it is possible to modify the curve. Such a representation is perfectly suited to represent linear segments as to a series of control points aligned in real space it immediately corresponds a straight segment of the corresponding curve. 

A B-spline of order $k$ is defined using a set of basis functions as
\begin{equation}
 \ds S(x) = \sum_{i} C_i\mathcal{B}_{i,k}(x), 
\end{equation}
where $C_i$ is the i-th control point of the curve and $\mathcal{B}_{i,k}$($x$) is the i-th \textit{basis function} of order k. A basis function of order $k$ will have non-zero values over $k+1$ knot intervals and so the lower is the degree of the basis functions, the more ``local" the effect of changing the coordinates of the respective control point. When crossing a knot point or equivalently when passing from a knot interval to the next, one of the control points stops influencing the next section of the curve while a new control point takes its place. The values of the basis functions of order $k$ can be computed iteratively using the Cox-De Boor algorithm \cite{de_boor}:
\begin{align}
\mathcal{B}_{i,0}(x) & = \begin{cases} 1, \hspace{2em} t_i < x < t_{i+1} \\ 0, \hspace{2em}  otherwise \end{cases} \\
\mathcal{B}_{i,k}(x) & = \frac{x-t_i}{t_{i+k}-t_i}\mathcal{B}_{i,k-1}(x) + 
\frac{t_{i+k+1} - x}{t_{i+k+1}-t_{i+1}}\mathcal{B}_{i+1,k-1}(x) ,
\end{align}
where $t_i$ is an element of the knot vector $(t_1,t_2...t_m)$ in which the knot points have been placed in ascending order.

\begin{figure}
 \centering
\includegraphics[width=0.5\textwidth]{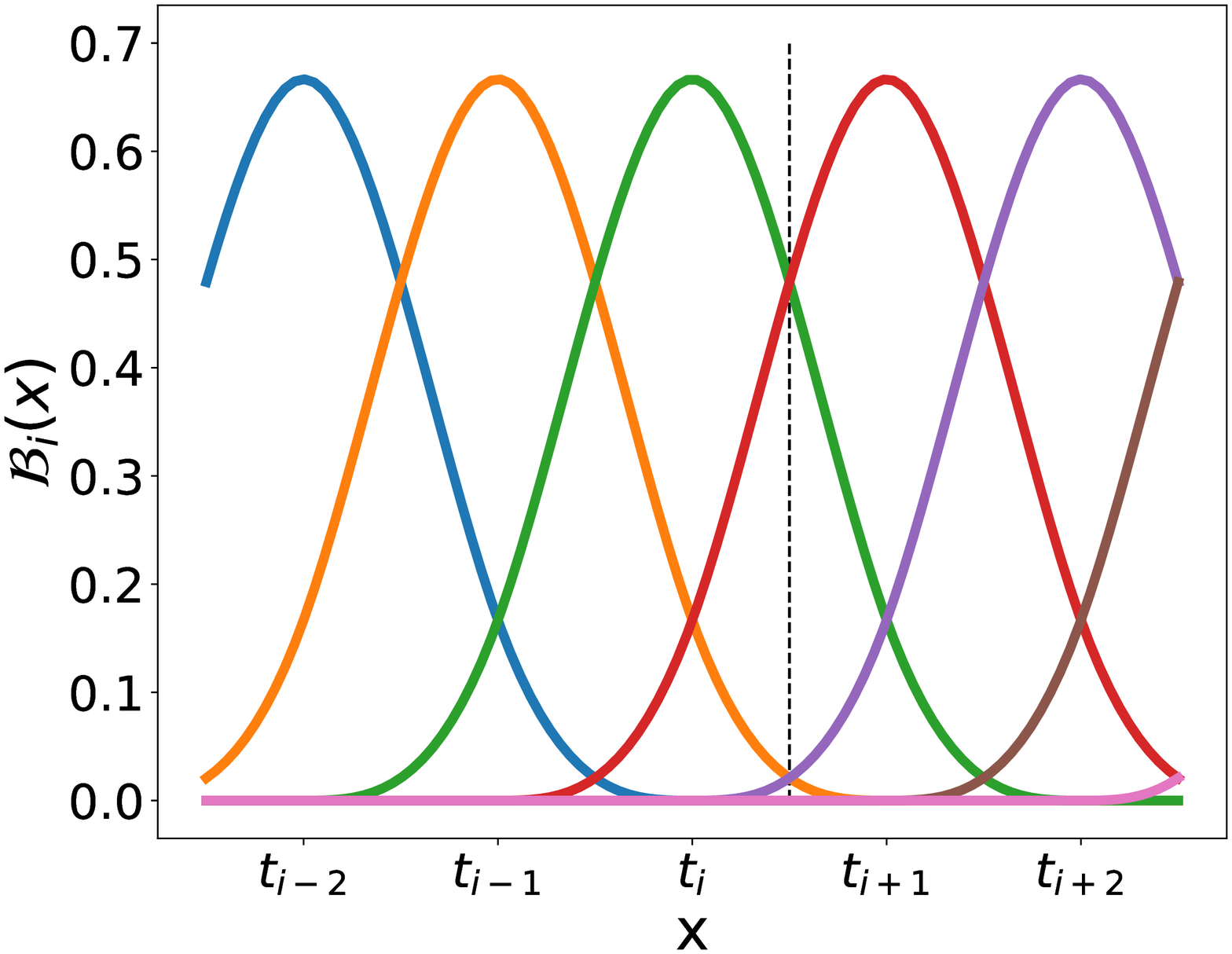} 
\caption{Third order basis functions corresponding to different control points. For every point inside a knot interval (e.g. the dotted line) there are only four basis functions with non-zero value.}
\label{fig0}
\end{figure}

In general a B-spline defined with $n$ control points requires
$n+k+1$ knot points in the knot vector ($t_1,t_2...t_{n+k+1}$) but for a closed curve the first $k$ control points must be overlapping with the last $k$ ones. 
In FOCUS third order (cubic) splines have been implemented and a plot of the third order basis functions corresponding to different control points are shown in figure \ref{fig0}.
It is worth noting how the values of the basis functions at every point x always sum up to 1 and that every basis function is non-zero only over 4 contiguous knot intervals.  
Cubic splines are a standard in the computer graphics field\cite{graph_gems} as it is the lowest order that is still able to have a flex point and in general lower order curves are preferred as they are more stable to oscillations due to a change in the control points coordinates.
During the optimization the derivatives of the coordinates with respect to the parameter $x$ are required, so the derivatives of the cubic basis functions were analytically computed and implemented into the code. The optimization is performed on the coordinates of the control points, while leaving the points used in the discretization of the curve at a constant value of the parameter $x$ (with $x$ $\in [t_1,t_{n+k+1}]$ ). Evaluating the points along the curve always at the same values of $x$ in  parameter space allows to compute the basis functions and their derivatives only once, during the initialization, and to treat them as constants during the execution.

\section{New cost function straightening coils and optimization procedures}\label{sec:opt}
Starting from an initial guess for the coils in B-spline representation (that can be obtained by manually manipulating the control points or fitting to known shapes), the code optimizes the coordinates of the control points and the current in each coil using optimization algorithms. The target function to be minimized consists of different cost functions that are multiplied by user-defined parameters called \textit{weights} and summed. 
The first-order derivatives are computed analytically. 

\subsection{Existing FOCUS cost functions used in this work}
FOCUS has several cost functions covering both physics requirements and engineering constraints.
All of the cost functions (except the one associated with the Fourier representation) have been reformed to use the spline representation.
While they all have specific purposes, we mainly use some cost functions in this work.
In particular, the relevant cost functions are:

\begin{itemize}
\item \textbf{Normal field error}
The following cost function implements the main physical requirement for the coils: generating a target magnetic field. In particular, FOCUS will try to minimize the difference between the normal field generated by the coils and a target normal field on the boundary. This can be expressed as:
\begin{equation}
f_B = \int_S \frac{1}{2}(\bi{B}_{coils} \cdot \bi{n} - B_n^{target})^2 \dd{a} ,
\end{equation}
where $\textbf{B}_{coils}$ is the field generated by the coils, $\vect{n}$ the normal vector of the boundary and $B_n^{target}$ the target normal magnetic field assessed from equilibrium codes.
For example, if we choose the last-closed-flux-surface as the boundary on which the total normal field is zero, we have $B_n^{target} = - \vect{B}_{plasma} \cdot \vect{n}$.

\item \textbf{Length}
To avoid trivial solutions where the coils tend to move away from the plasma to reduce the field at the surface, a constraint on the length is present in FOCUS. For the optimizations in this work the quadratic form for the length penalty has been used
\begin{equation}
f_L = \frac{1}{N_c} \sum_{i=1}^{N_c} \frac{1}{2}\frac{(L_i-L_{i,o})^2}{L_{i,o}^2} ,
\end{equation}
where $N_c$ is the number of coils, $L_{i,o}$ is the target length and $L_i$ is the length of the $i$-th coil.

\item \textbf{Curvature}
To design coils that are more easily manufacturable, a constraint to reduce the curvature of the coils is present. While different forms have been implemented in FOCUS \cite{kruger_zhu_bader_anderson_singh_2021}, the one used for the optimizations in this work is the constrained curvature penalty given by:
   \begin{equation}
    f_{\kappa} = \frac{1}{N_c}\sum_{i=1}^{N_c} \int_0^{2\pi} H_{\kappa_o}(\kappa_i) (\kappa_i - \kappa_o)^\alpha dt ,
    \end{equation}
where $H_{\kappa_o}(\kappa_i)$ is the step function,  $\kappa_o$ and $\alpha$ are user-defined parameters and $\kappa_i$ is the curvature at a point defined as $\kappa_i = \lVert \vect{x}' \times \vect{x}'' \rVert / \lVert \vect{x}''' \rVert $.
$\kappa_o$ has the meaning of a maximum allowed curvature and the cost function has no effect on coil segments with a curvature smaller than $\kappa_o$.
\end{itemize}

\subsection{Cost function for straight \change{coils}{sections}}
The locality of the spline representation was used to constraint the design to favor straighter outer coils. 
In particular, a cost function with the same type of penalty as the curvature cost function but applied only to a subset of the points making up the curve was implemented.
For example, considering the constrained curvature objective function, the new cost function is given by
    \begin{equation}
    f_{S} = \\ \frac{1}{N_c}\sum_{i=1}^{N_c} \int_0^{2\pi} W(t)H_{\kappa_{s,0}}(\kappa_{i}) (\kappa_{i} - \kappa_{s,o})^{\alpha_s} dt \ ,
    \end{equation}
where $W(t)$ is a weight function, $H_{\kappa_{s,0}}$ is the step function and $\kappa_{s,o}$ and $\alpha_s$ are user-defined parameters.
In this case $\kappa_{s,o}$ is the maximum curvature allowed on the outboard side while $\alpha_s$ determines the strength of the scaling. The weight function is defined as 
\begin{equation}
W(t) = \begin{cases} 1, \hspace{2em} P_{xy}(t) > P_m + \beta P_d \\ 0, \hspace{2em}  otherwise \end{cases}
\end{equation}
where $P_m$ is the mean squared distance of the coil projection in the x-y plane from a user-defined point and $P_d$ is a measure for half the maximum projection in the x-y plane of the coil radius. Considering the distance of the projection of a point along the coil in the x-y plane from the user-defined point ($x_0,y_0$)
\begin{equation}
P_{xy}(t) = (x_i(t) - x_0)^2 + (y_i(t) - y_0)^2 \ ,
\end{equation}
they are defined as 
\begin{align}
& P_m = \overline{P_{xy}} \ , & P_d = \frac{\max(P_{xy}) - P_m}{2} \ .
\end{align}
$\beta$ is a user-defined parameter that can be used to tweak the section of the coil affected by the optimization. With a value $\beta = 0$ the section of the coil with a projected distance greater than the mean value for that coil will all be affected while for a value $\beta = 2$ the new cost function will not be applied to any point.
While it may be tempting to use the three-dimensional distance to determine the outer section of the coil instead of the projection along the x-y plane, such a metric is affected by oscillations in the z-direction which may lead to disjointed sections of the coil begin affected and so this metric has been adopted instead.      

\subsection{Quick implementation of analytical derivatives using functional derivatives}
For general optimization problems, it is useful to have the information about the derivatives, especially when one wants to use gradient-based optimization algorithms.
New methods, like the adjoint method \cite{Paul2018} and automatic differentiation \cite{McGreivy2021}, have been brought into the field of stellarator optimization.
Here, we introduce a different approach using the functional derivative.

The derivation of functional derivatives for FOCUS cost functions was initially introduced in \cite{Zhu_2018}. Later, it was extended to the variation of surface in \cite{Hudson2018}. Here, we will briefly revisit the derivation using the cost function $f_B$ as an example.
The magnetic field generated by the coils is calculated using the Biot-Savart law,
\be
\vect{B}_{coils} = \ds \sum_{i=1}^{N_c} \vect{B}_i = \ds \sum_{i=1}^{N_c} \frac{\mu_0 I_i}{4\pi} \int_C \frac{\dd{\vect{l}}_i \times \vect{r}}{r^3} \ ,
\ee
where $\dd{\vect{l}}_i = \vect{x}_i' \dd{t}$, $\vect{r}=\bar{\vect{x}} - \vect{x}_i$ ($\vect{x}_i$ a point on the $i$-th coil, $\bar{\vect{x}}$ the evaluation point on the surface).
$f_B$ is a functional of the coil shapes.
If we vary the geometry of the $i$-th coil, the first variation is
\be
\delta f_B = \ds \int_S (\bi{B}_{coils} \cdot \bi{n} - B_n^{target}) \ \delta \vect{B}_i \cdot \vect{n} \dd{a} \ .
\ee
The variation of the magnetic field is calculated as
\be
\delta \vect{B}_i = \int_C (\delta \vect{x}_i \times \vect{x}'_i) \cdot \vect{R}_i \dd{t} \ ,
\ee
where $\vect{R} = 3 \vect{r} \vect{r}/r^5 - \vect{I}/r^3$ ($\vect{I}$ is a unit dyadic). So far, all the equations are general and we haven't touched any specific coil parameterization. To calculate the derivatives of $f_B$ with respect to the degrees of freedom (\eg Fourier coefficients in the Fourier representation or control points in the spline representation), we can simply apply the chain rule,
\be
\pdv{f_B}{\vect{X}} = \frac{\delta f_B}{\delta \vect{x}} \pdv{\vect{x}}{\vect{X}} \ . \label{eq:fderiv}
\ee
Here, $\vect{X}$ denotes the coil parameters.
The first term on the right-hand side of \Eqn{fderiv} is parameterization-free and the second term is parameterization-specific. When we change the coil representation, we only need to modify the second term, which is usually simple.
For example, in this work, the derivative of the coil shape with respect to the position of a control point is
\be
\pdv{\vect{x}}{C_i} = \mathcal{B}_{i,k} \ .
\ee

Normally, one can directly differentiate functions and write down the derivatives with respect to parameters.
This process, which is usually called ``symbolic differentiation'', will compute derivatives accurately and fast.
However, for complex functions, like the ones we used in this paper, the derivatives have many terms and are thus difficult to be numerically implemented.
When changing to a new parameterization, one has to derive the equations and implement the derivatives one more time, which requires non-trivial effort.
In this paper, we split the shape part and the parameter part using functional derivatives, like in \Eqn{fderiv}.
By doing so, when changing to a new parameterization, the only parts that have to be re-implemented are the derivatives of the coil shape with respect to the coil parameters.
The method will be extremely useful for situations with multiple parameterizations (for curves, surfaces, and other geometries).

\section{Numerical applications}\label{sec:applications}
Now, we can use the new code to design coils with improved access on the outboard side.
In the following, the new representation and cost function were applied to design an improved set of coils for the Helically Symmetric eXperiment (HSX) and the National Compact Stellarator Experiment (NCSX)  stellarators.
Furthermore, we are going to design a new set of coils for a prototype quasi-axisymmetric stellarator (QAS) reactor, which has considerably large outboard access.
Due to the toroidal periodicity and stellarator symmetry, FOCUS only needs to design a set of coils in a half period and the remaining coils are then obtained by reflection and repetition of the designed set.

Tests using a variety of weights were performed, studying the variation of the surface-normalized normal field error, $<{B_n}> = \int \abs{B_n}/{B}\dd{S}/{A}$, which is taken as the main figure of merit in the rest of the discussion along with the reduction of the curvature on the outboard side.
The optimized coil sets for the first two experiments were obtained by trying to minimize the curvature on the outboard side while still maintaining the same maximum curvature along the entire coil present in the coils of the real experiment and a $<B_n>$ lower than the value corresponding to the real coils.
For the QAS reactor-like configuration, the quality of the coils was determined by how closely the flux surfaces resembled the target boundary.
The best results found by exploring the parameter space are presented in the rest of the section. 

\subsection{HSX}
HSX \cite{hsx} is a stellarator optimized for quasi-helical symmetry built at the University of Wisconsin-Madison. It has a major radius of 1.2 m with 4 field periods and 12 magnets per period divided in 2 sets of 6 magnets related by the stellarator symmetry. In this section, the coils obtained using the spline representation are compared with the coils presented in \cite{kruger_zhu_bader_anderson_singh_2021}, obtained using the constrained curvature function and already shown to be an improvement over the actual coils.

\begin{figure*}
\centering
\includegraphics[width=0.75\textwidth]{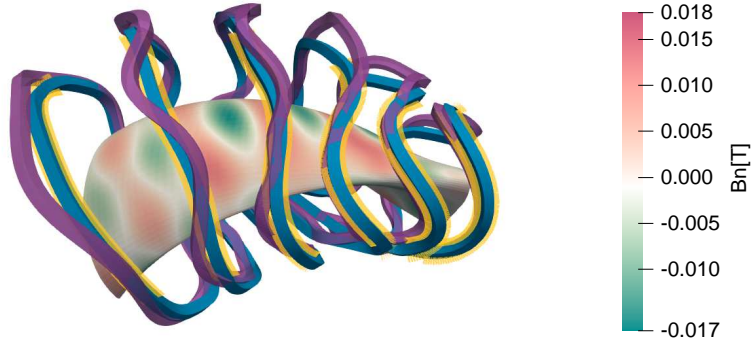} \caption{Comparison of the coils designed with splines for HSX (blue) and the ones shown in \cite{kruger_zhu_bader_anderson_singh_2021} (violet). The golden area of the coil is the part affected by the straightening cost function. The color of the surface represents the $B_n$ at that position.}
\label{fig1}
\end{figure*}

In figure \ref{fig1} a 3D representation of the plasma surface and the two coil sets is shown.
FOCUS optimizes the coils as one-dimensional filaments but they are shown as finite-build coils to have a more realistic view of what they would look like once built.
A larger straightening effect can be seen on the second and third coils from the left in the front view.
Some characteristics of this coil set are compared with the reference one in table \ref{tab:diagnostics} and it can be noticed that this configuration is able to have a straighter section on the outside while still providing a reasonable $<B_n>$.
Here, we are not trying to match $<B_n>$ from the reference coil set, instead we stopped when $<B_n>$ was lower than the value of the real coils ($<B_n> = 4.87 \times 10^{-3}$).
In figure \ref{fig4} a comparison of the flux surfaces obtained from field line tracing is shown and the new configuration can follow the target boundary as well as the reference configuration.
In appendix \hyperref[sec:HSX_field_QH]{A}, the quality of the magnetic field for the new coil-set is compared to the reference one to show that the slight increase of $<B_n>$ does not affect significantly the quasi-helical symmetry. 

\begin{table}
\centering
\caption{Comparison of reference and optimized coils characteristics for HSX}
\label{tab:diagnostics}
\begin{tabular}{lll}
\toprule
Coil Set & Ref. & Opt.\\
\midrule
Maximum curvature & 12.3 m$^{-1}$ & 12.3 m$^{-1}$ \\
Average curvature & 6.61 m$^{-1}$ & 4.39 m$^{-1}$ \\
Average Length & 2.38 m &   2.23 m\\
$<B_n>$ & 1.15$\times 10^{-3}$ & 4.52$\times 10^{-3}$ \\
Minimum coil-coil \\distance on outboard side & 0.076 m& 0.090 m\\
\bottomrule
\end{tabular}

\end{table}

\begin{table}
\centering
\caption{Comparison of outboard side mean and max curvature $\kappa$ [m$^{-1}$]  for HSX}
\label{tab:curve}
\begin{tabular}{lllll}
\toprule
Coil & Ref. $\overline{\kappa}$ & Opt. $\overline{\kappa}$ & Ref. max $\kappa$& Opt. max $\kappa$  \\
\midrule
1& 3.27  & 2.57  & 4.53 & 3.26  \\
2& 5.15 & 2.42 & 12.02  & 3.17  \\
3& 6.78  & 3.24  & 12.27  & 5.9\\
4& 7.07  & 4.76  & 12.29  & 6.79 \\
5& 6.44  & 4.67  & 12.24  & 6.98\\
6& 5.59  & 4.97  & 8.68  & 6.00 \\

\bottomrule
\end{tabular}

\end{table}

\begin{figure*}
\centering
\includegraphics[width=0.32\textwidth]{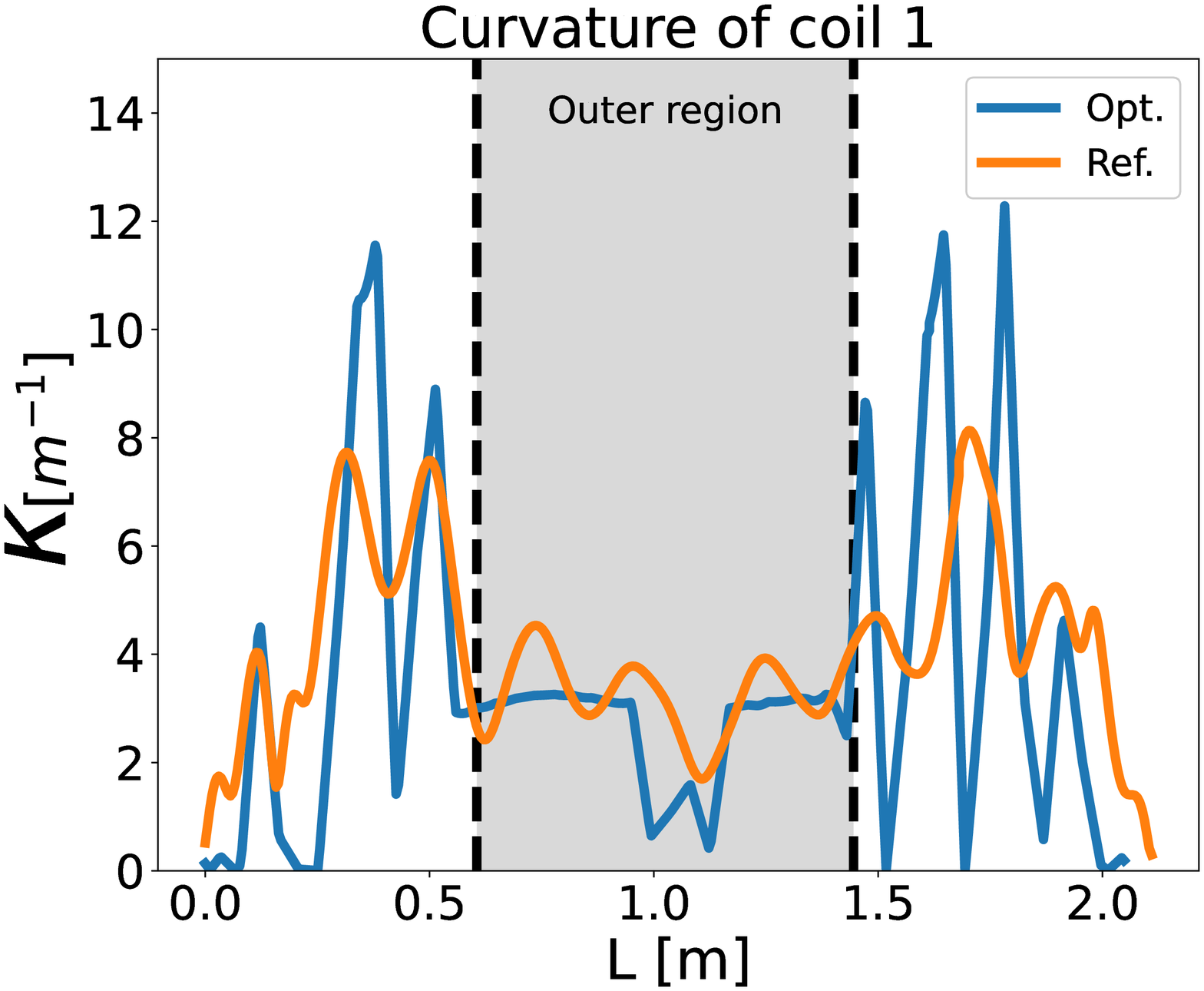}
\includegraphics[width=0.32\textwidth]{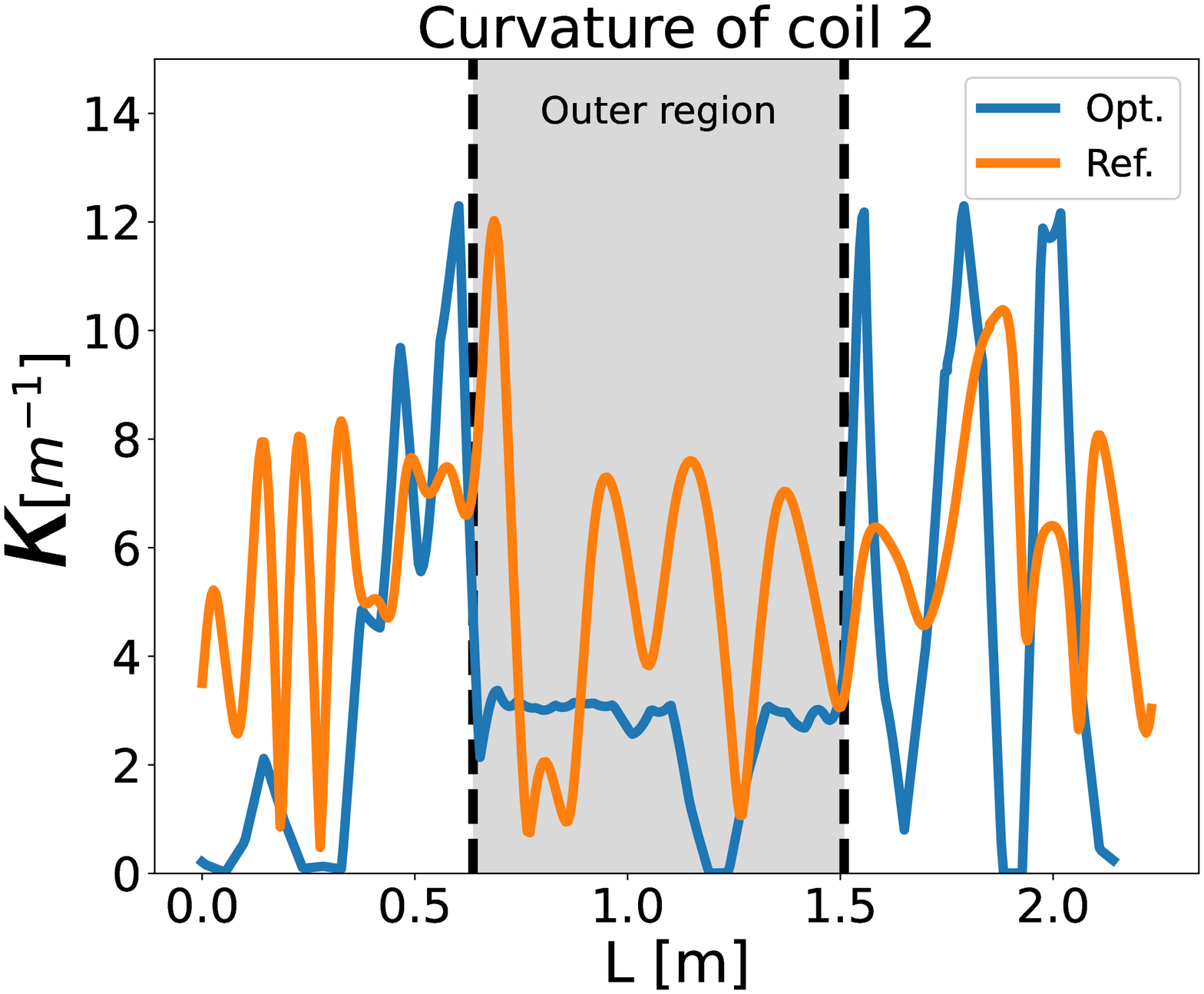}
\includegraphics[width=0.32\textwidth]{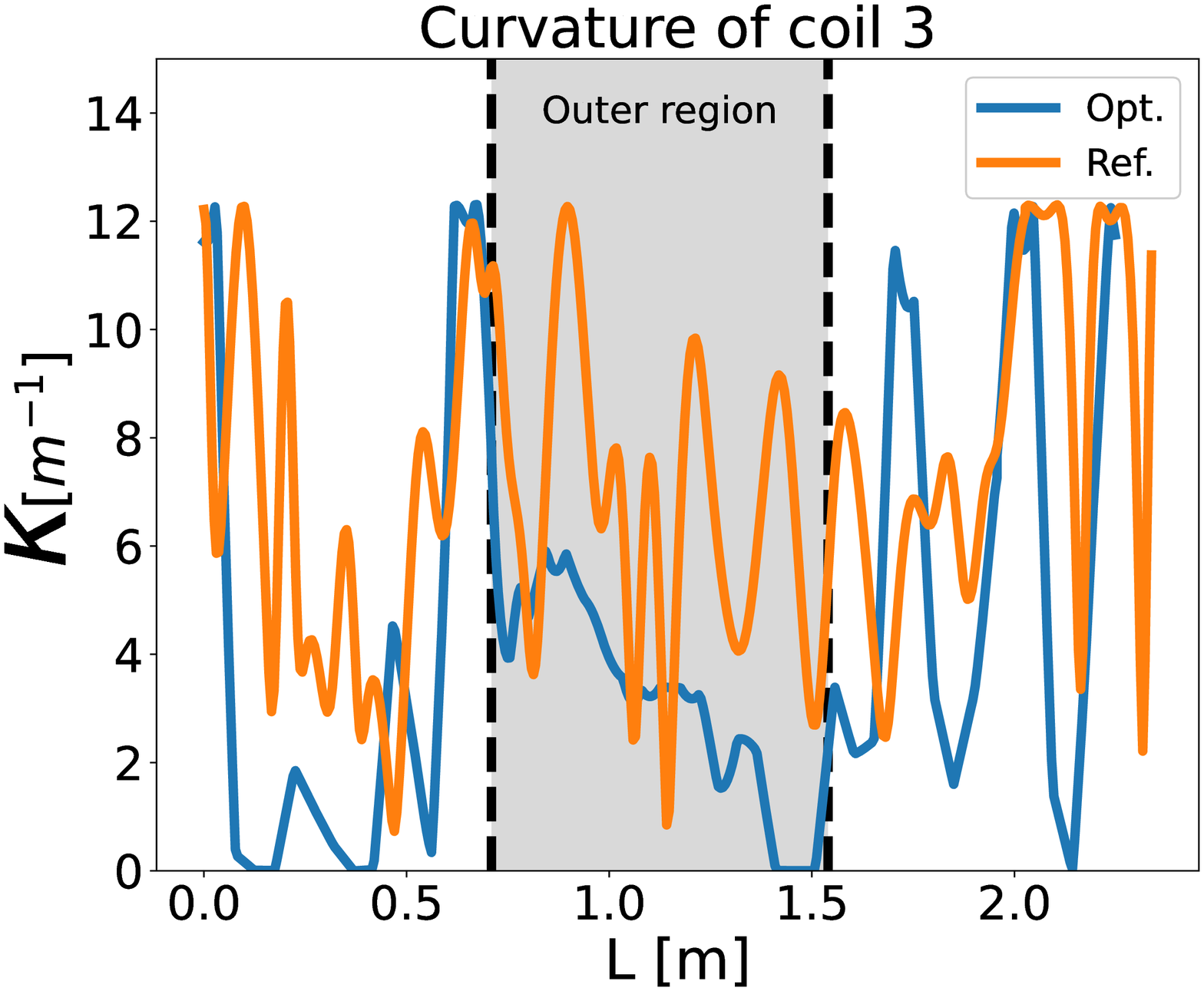}
\includegraphics[width=0.32\textwidth]{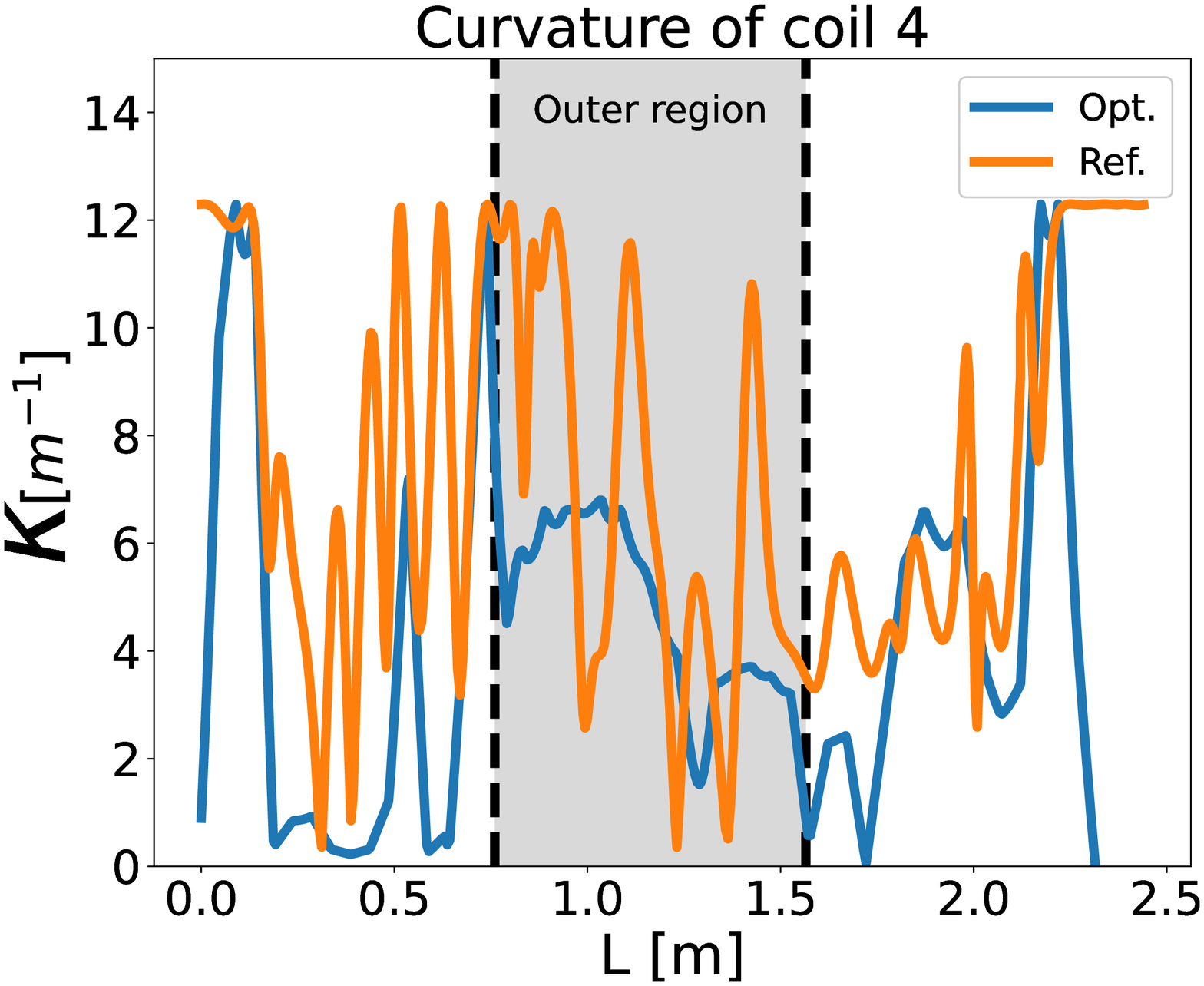}
\includegraphics[width=0.32\textwidth]{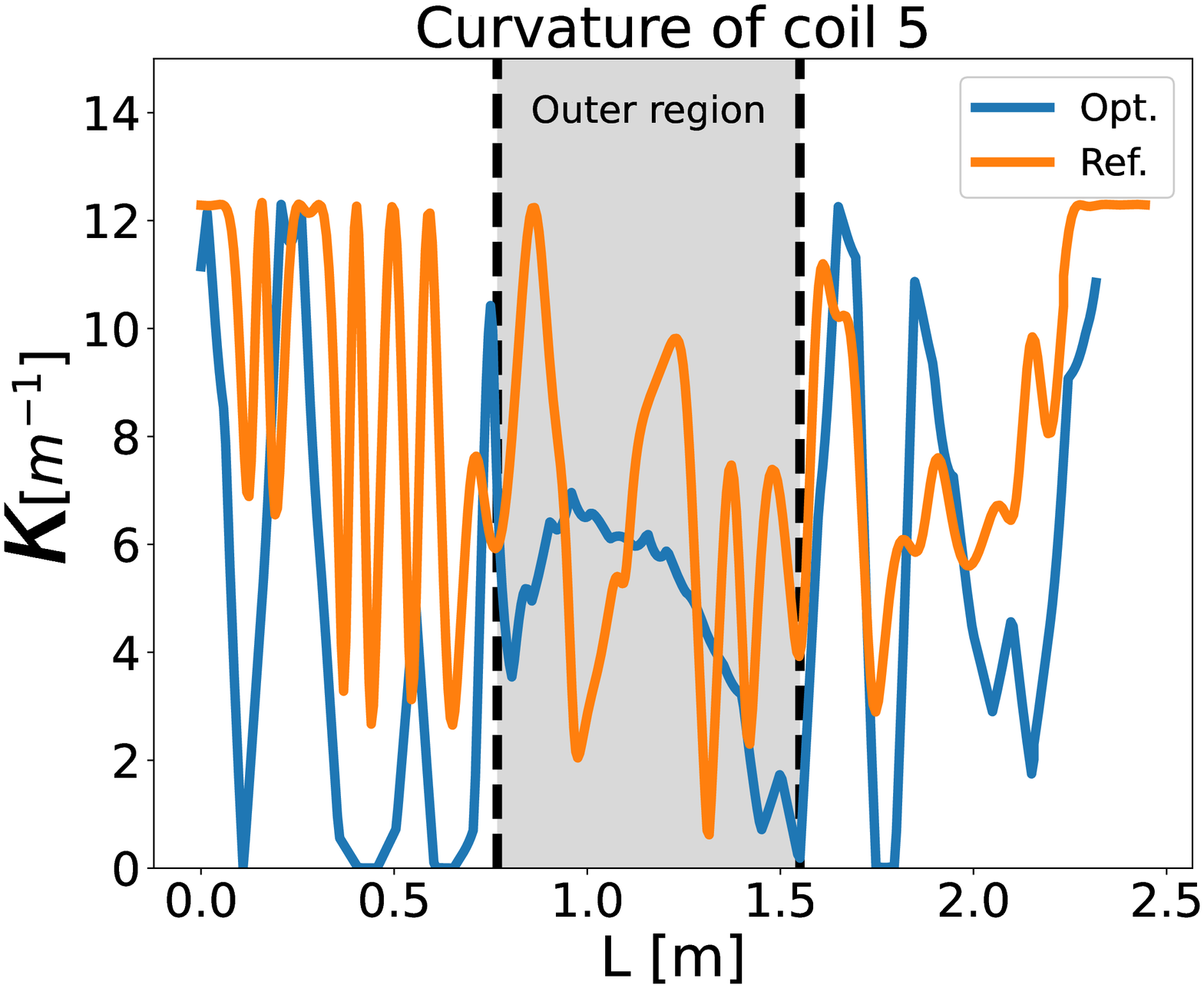}
\includegraphics[width=0.32\textwidth]{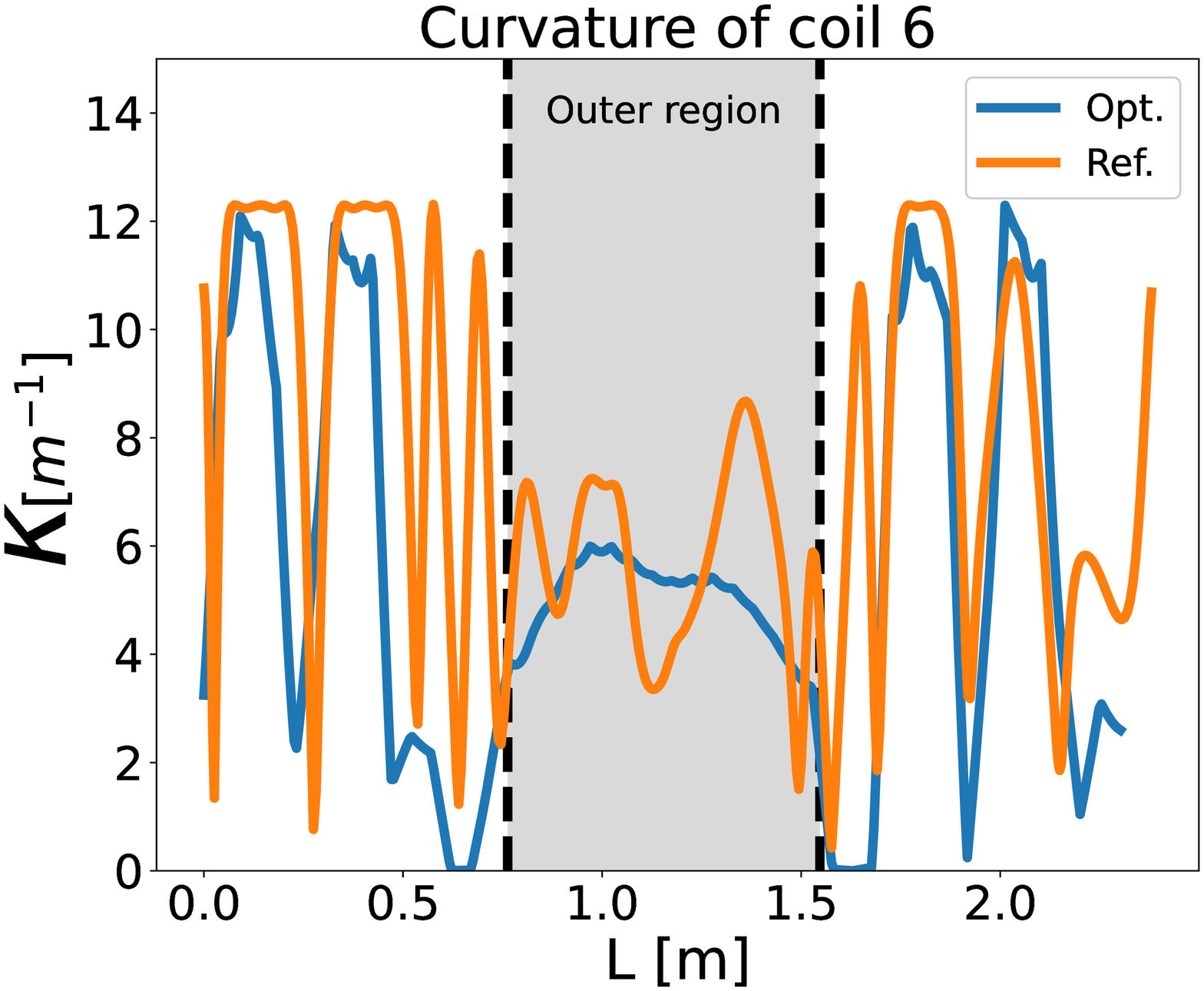}
\caption{Comparison of the curvature $\kappa$ for each coil in HSX along the length of the coil L. The gray areas indicate the regions of the optimized coils affected by the straight-out cost function. Coil 1 represents the left-most coil in Figure \ref{fig1} and coil 6 the right-most coil.}
\label{fig2}
\end{figure*}

\begin{figure}
\centering
\includegraphics[width=0.5\textwidth]{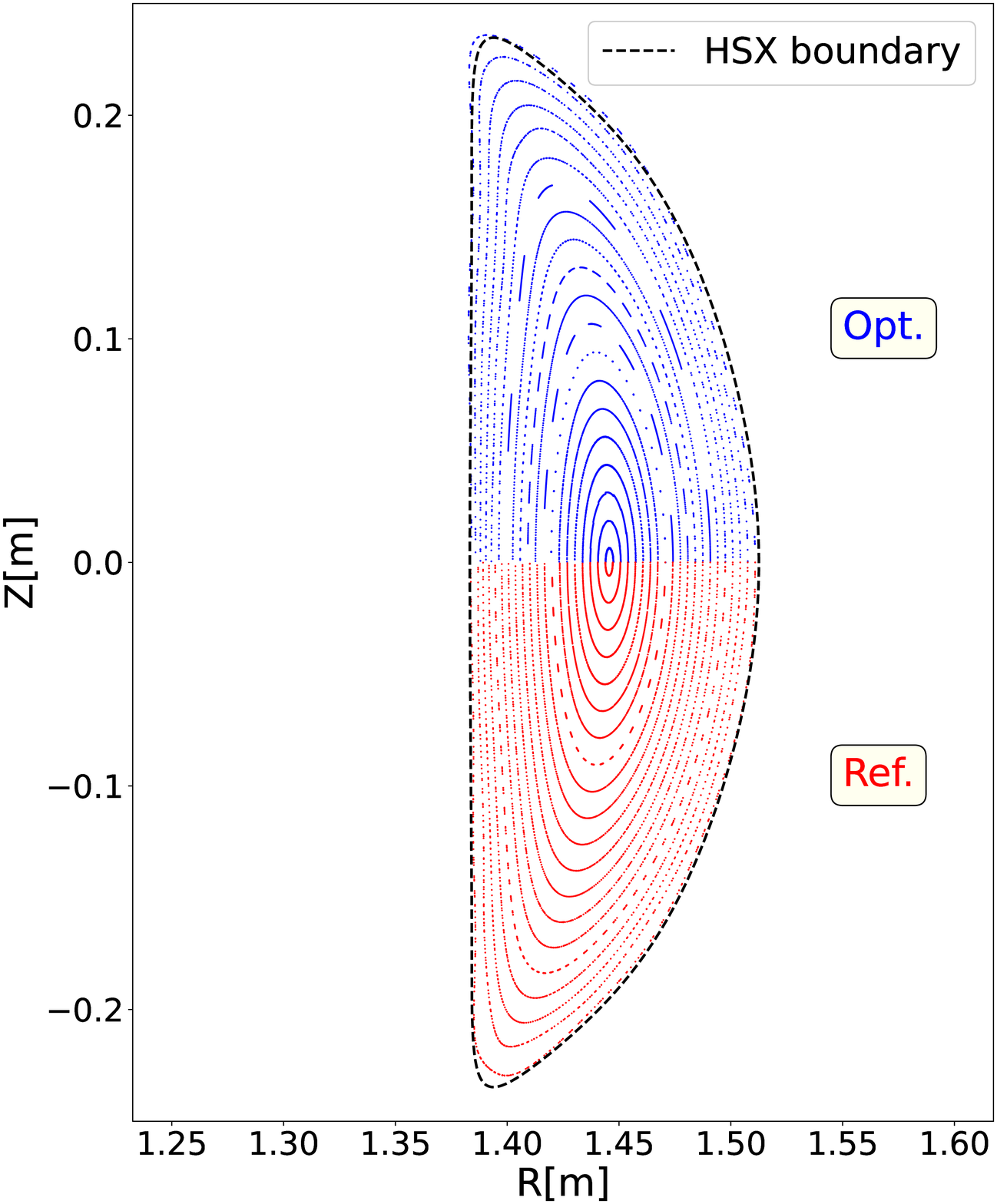} 
\caption{Comparison between the Poincarè plot obtained from the coils in \cite{kruger_zhu_bader_anderson_singh_2021}(red) and the optimized coils (blue) for $\phi$ = 0 in HSX using the FOCUS field line tracing algorithm.}
\label{fig4}
\end{figure}

In figure \ref{fig2} the curvature along each coil is compared with the reference set. The improvement on the outer part is evident and the values of the mean and maximum curvature in the outer regions are compared in table \ref{tab:curve}. The optimized configuration has been obtained using the same $\kappa_0 = 12.3$ m$^{-1}$ of the real coils but restricting the maximum curvature on the outer side to be  $\kappa_s = 3$ m$^{-1}$ and using $\beta = 0.5$

\subsection{NCSX}
\begin{figure}
\centering

\includegraphics[width=\linewidth]{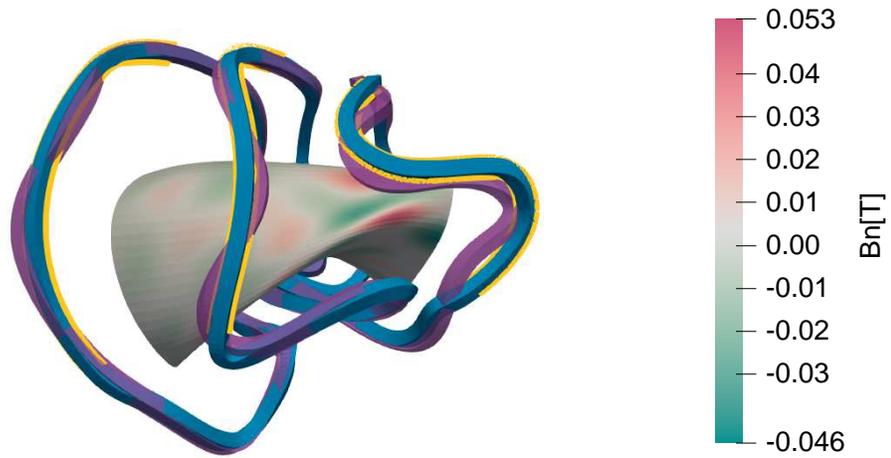} 
\caption{Comparison of the coils designed with splines (blue) and the ones designed originally for the NCSX experiment (violet). The golden area of the coil is the part affected by the straightening cost function. The color of the surface represents the $B_n$ at that position.}
\label{fig7}

\end{figure}

\begin{figure}
\centering
\includegraphics[width=0.32\textwidth]{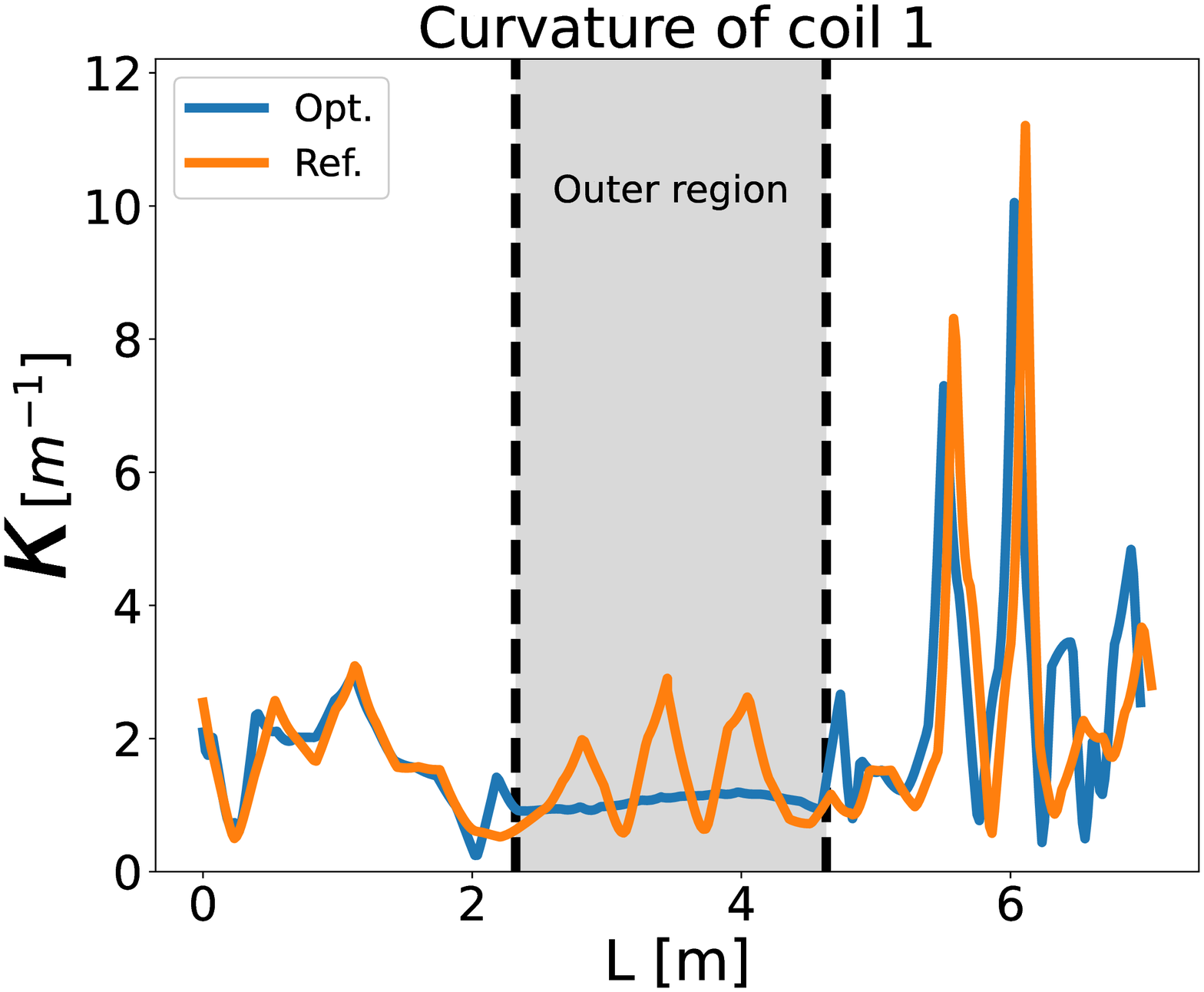}
\includegraphics[width=0.32\textwidth]{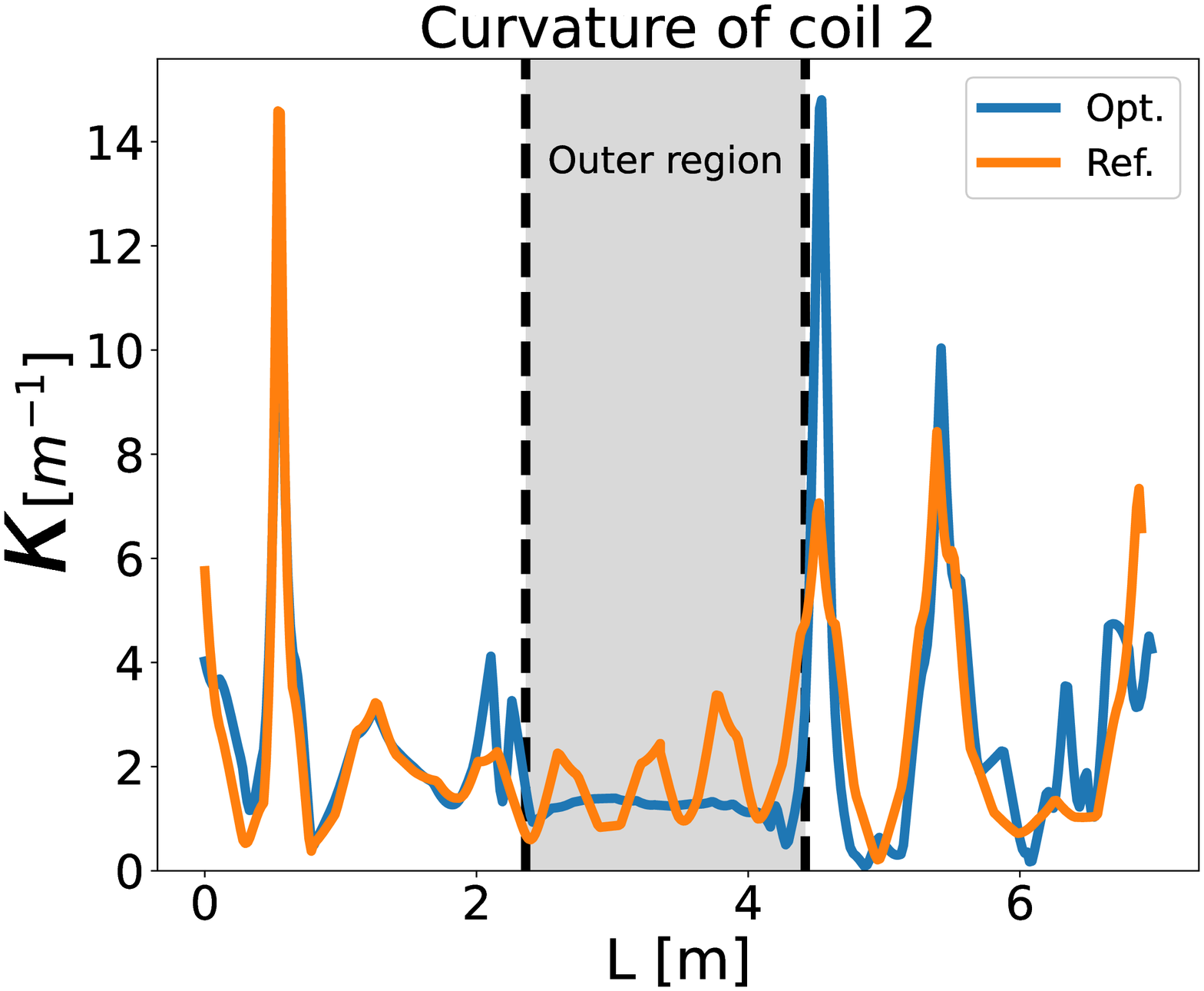}
\includegraphics[width=0.32\textwidth]{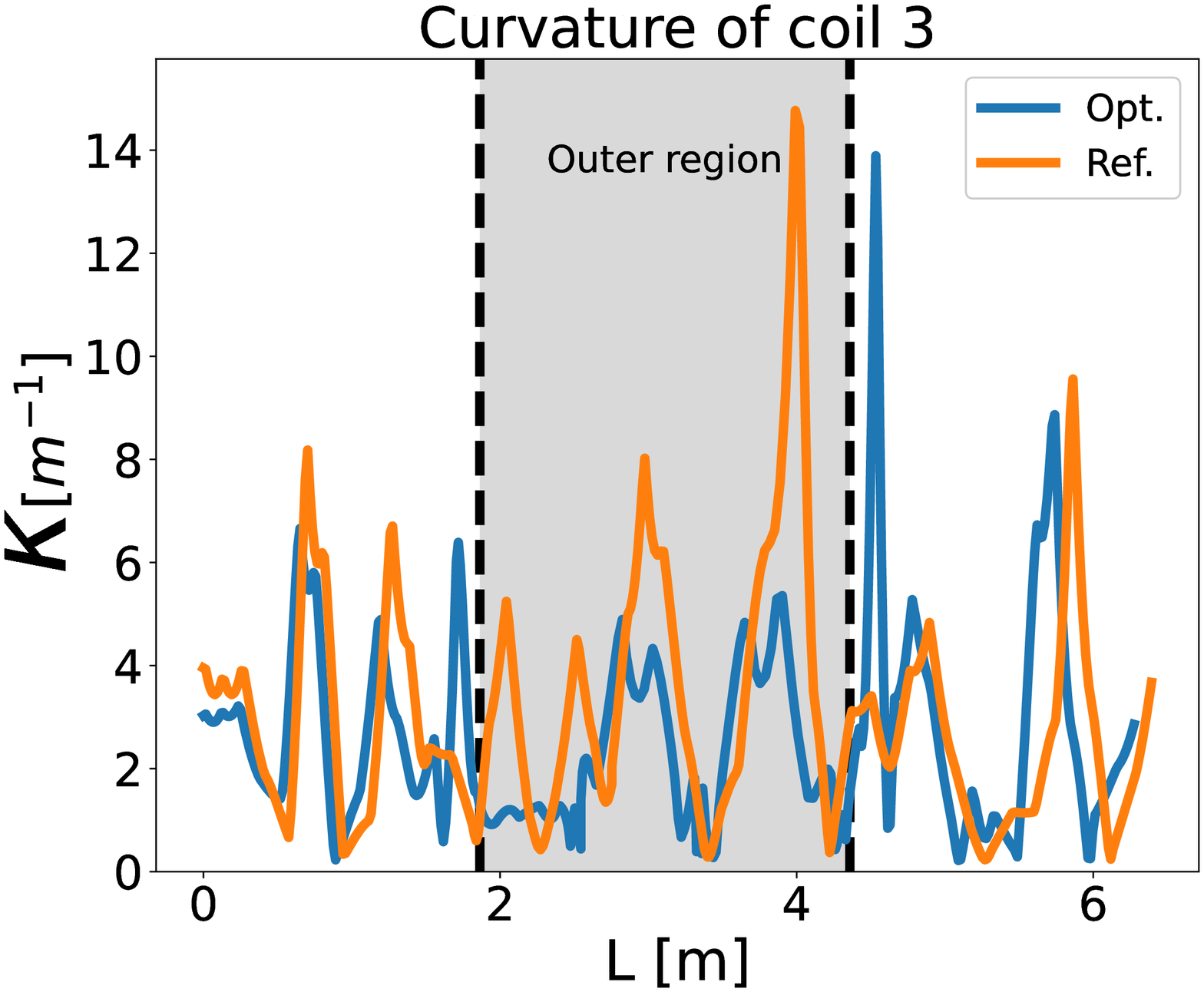}
\caption{Comparison of the curvature $\kappa$ for each coil in NCSX. The gray areas indicate the regions of the optimized coils affected by the straight-out cost function. Coil 1 represents the left-most coil in Figure \ref{fig7} and coil 3 the right-most coil.}
\label{fig8}
\end{figure}

\begin{figure}
\centering
\includegraphics[width=0.5\textwidth]{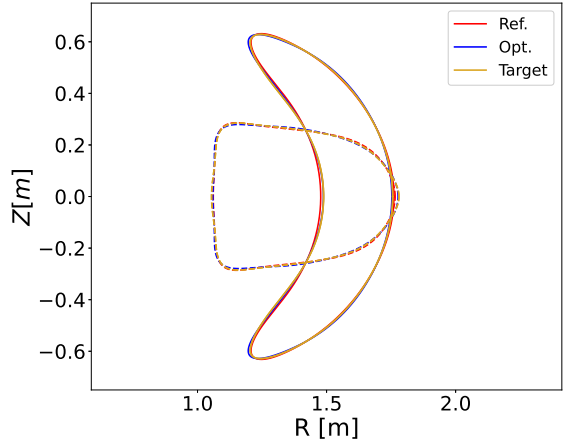} 
\caption{Comparison between the VMEC solutions obtained from the coils initially designed for the experiment (red) and the optimized coils (blue) for $\phi$ = 0 (full lines) and $\phi = \frac{\pi}{3}$ (dotted lines) in NCSX.}
\label{fig9}
\end{figure}

NCSX is a quasi-axisymmetric stellarator that was partly built at the Princeton Plasma Physics Laboratory but was unfortunately canceled \cite{Neilson2010}.
The target equilibrium, labeled as C09R00, has a major radius of 1.44 m and a minor radius of 0.32 m.
It has 3 toroidal field periods and the main coil system consists of 12 modular coils in 3 unique shapes.
The new spline representation has also been applied successfully to improve the design for NCSX. In figure \ref{fig7} the improved coils are compared to the ones envisioned for the experiment and the improvement can be noticed also by looking at the curvature plots for the coils shown in figure \ref{fig8}.
In figure \ref{fig9} the flux surfaces obtained using the free-boundary VMEC \cite{VMEC_fb} with the two coil sets are compared with the fixed-boundary target.
As expected from similar $<B_n>$ values, the two configurations can follow the target surface similarly well. 
In this case, the VMEC code has been used to include the effect of the non-zero target normal field $T_{B_n}$ necessary to describe this plasma surface. 
The general characteristics of the two coil sets are compared in table \ref{tab:diag_NCSX} while their mean and average curvature of the outer section are compared in table \ref{tab:curve_NCSX}. While the minimum distance between two coils on the outer side is slightly lower in the optimized case, this minimum distance is reached at the edge of the region affected by the cost function. This upper ``junction point" between the inner and outer side of the coils lies above the plasma and so even if the coils were to be closer in that position, this would not affect the placement of ports around $z \approx 0$.

In this case, the optimization allowed for both a significantly simpler geometry as well as an improvement of the normal field error while leaving the inner side basically unmodified. 

 \begin{table}
\centering
\caption{Comparison of reference and optimized coils characteristics for NCSX}
\label{tab:diag_NCSX}
\begin{tabular}{lll}
\toprule
Coil Set & Ref. & Opt.\\
\midrule
Maximum curvature &  14.8 m$^{-1}$ & 14.8 m$^{-1}$ \\
Average curvature &  2.36 m$^{-1}$ &  2.28 m$^{-1}$ \\
Average Length &  6.81 m &  6.77  m\\
$<B_n>$  & 5.55 $\times 10^{-3}$ & 4.14 $\times 10^{-3}$ \\
Minimum coil-coil \\distance on outboard side & 0.29 m&  0.26 m\\
\bottomrule
\end{tabular}

\end{table}

\begin{table}
\centering
\caption{Comparison of outboard side mean and max curvature  $\kappa$ [m$^{-1}$] for NCSX.}
\label{tab:curve_NCSX}
\begin{tabular}{lllll}
\toprule
Coil & Ref. $\overline{\kappa}$ & Opt. $\overline{ \kappa}$& Ref. max $\kappa$ & Opt. max $\kappa$\\
\midrule
1& 1.38  & 1.05 & 2.90  & 1.19   \\
2& 1.81  & 1.23  & 4.74  & 2.16\\
3& 3.28  & 1.73  & 14.77  & 5.36 \\

\bottomrule
\end{tabular}
\end{table}

\subsection{QAS reactor}
Having verified the new capabilities of the code by comparing its results to known configurations, the coils designed by FOCUS for a prototype QAS reactor are presented in the rest of the section.
The target configuration comes from a study following \cite{Coilplot} and is obtained using a free-boundary STELLOPT optimization \cite{osti_1617636}.
Some main parameters are $N_{fp}=3$, $R=9.32$ m, $a=1.55$ m, $V=444.13 \mathrm{m}^3$, and $<B>=6.41$ T.

\begin{figure*}
\centering
\includegraphics[width=0.8\textwidth]{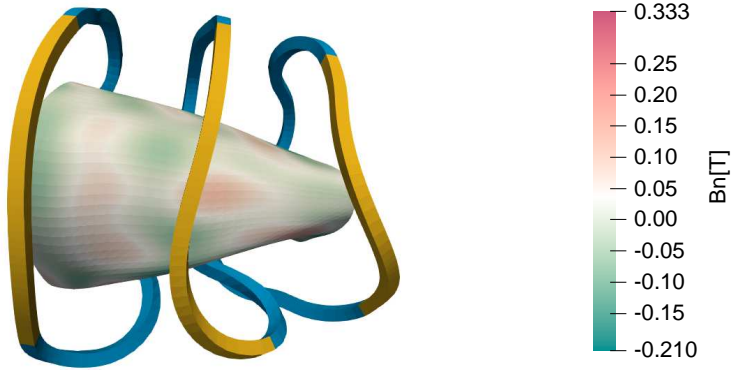} 
\caption{3D plot of the coils obtained by FOCUS for the prototype QAS reactor. The golden area of the coil is the part affected by the straightening cost function. The color of the surface represents the $B_n$ at that position.}
\label{fig10}
\end{figure*}

In figure \ref{fig10} a 3D plot of the coils and the plasma surface is shown.
The coils clearly appear to lack any rapid oscillations on the outer side. Most of the outer section is approximately straight and only curving back towards the bottom of the coils to join the inner side of the coils. 
This will be extremely suitable for remote access and blanket replacement.
In figure \ref{fig12} the curvature for every coil is plotted and its suppression in the outer region is evident. In the third coil, which corresponds to the region where the plasma twists at the end of a period, the larger value of the curvature in the outer region is due to the necessity of the coil to follow the curvature of the plasma. Increasing the strength of the constraint further leads to straighter coils also in this region but at the cost of having flux surfaces significantly different from the target boundary.

\begin{figure}
\centering
\includegraphics[width=0.5\textwidth]{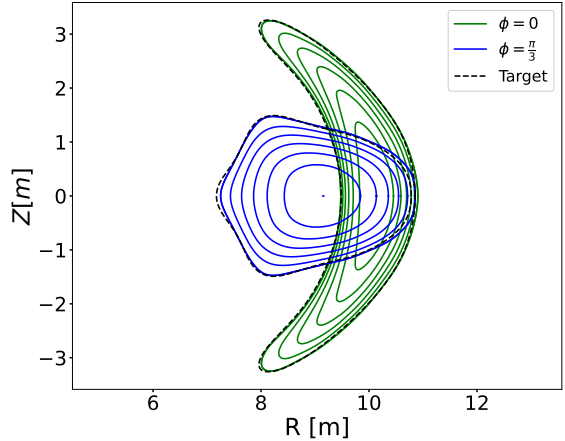} 
\caption{VMEC solutions for the coils designed using FOCUS for the prototype QAS reactor for $\phi = 0$ and $\phi = {\pi}/{3}$}
\label{fig11}
\end{figure}

\begin{figure}
\centering
\includegraphics[width=0.32\textwidth]{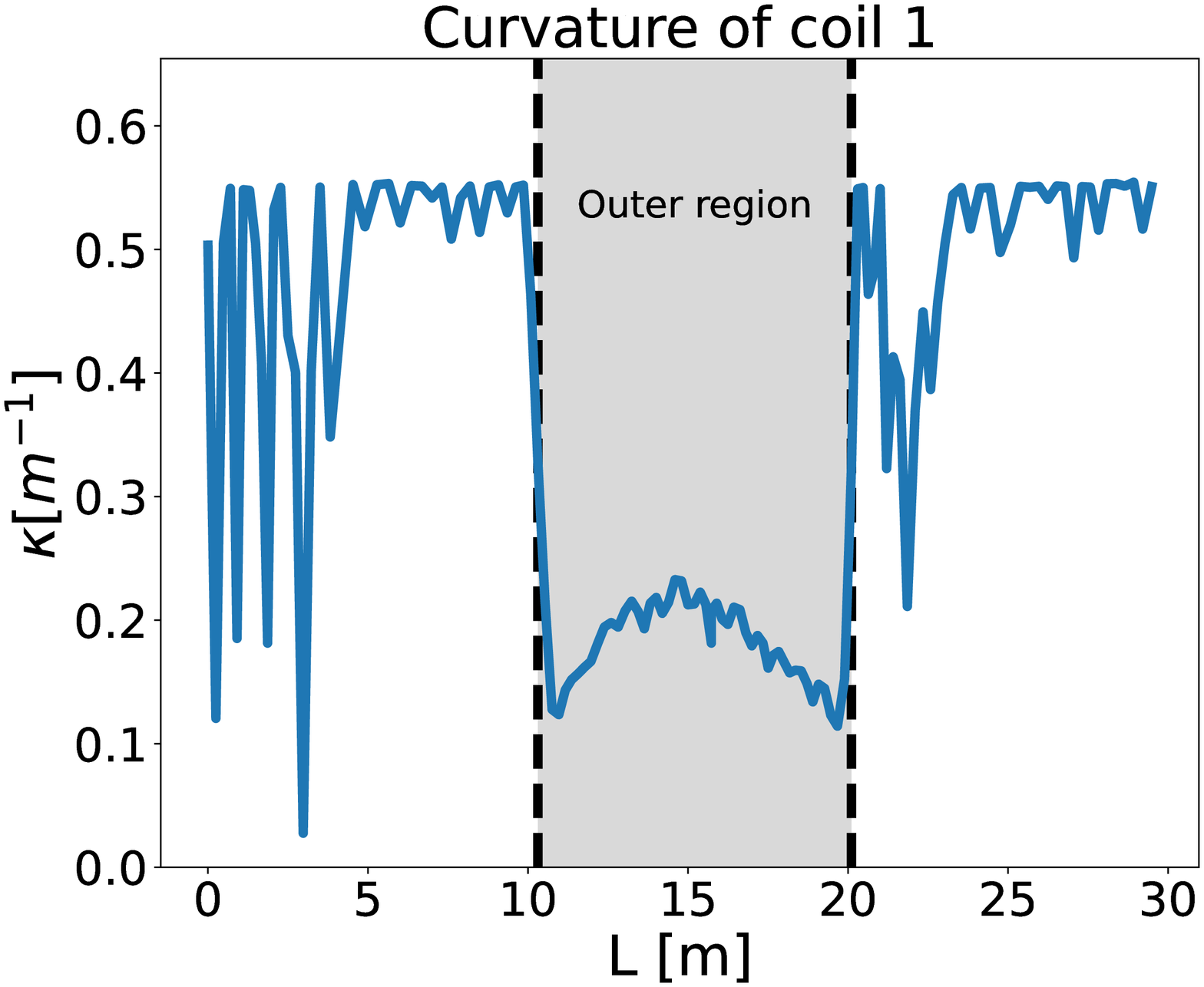}
\includegraphics[width=0.32\textwidth]{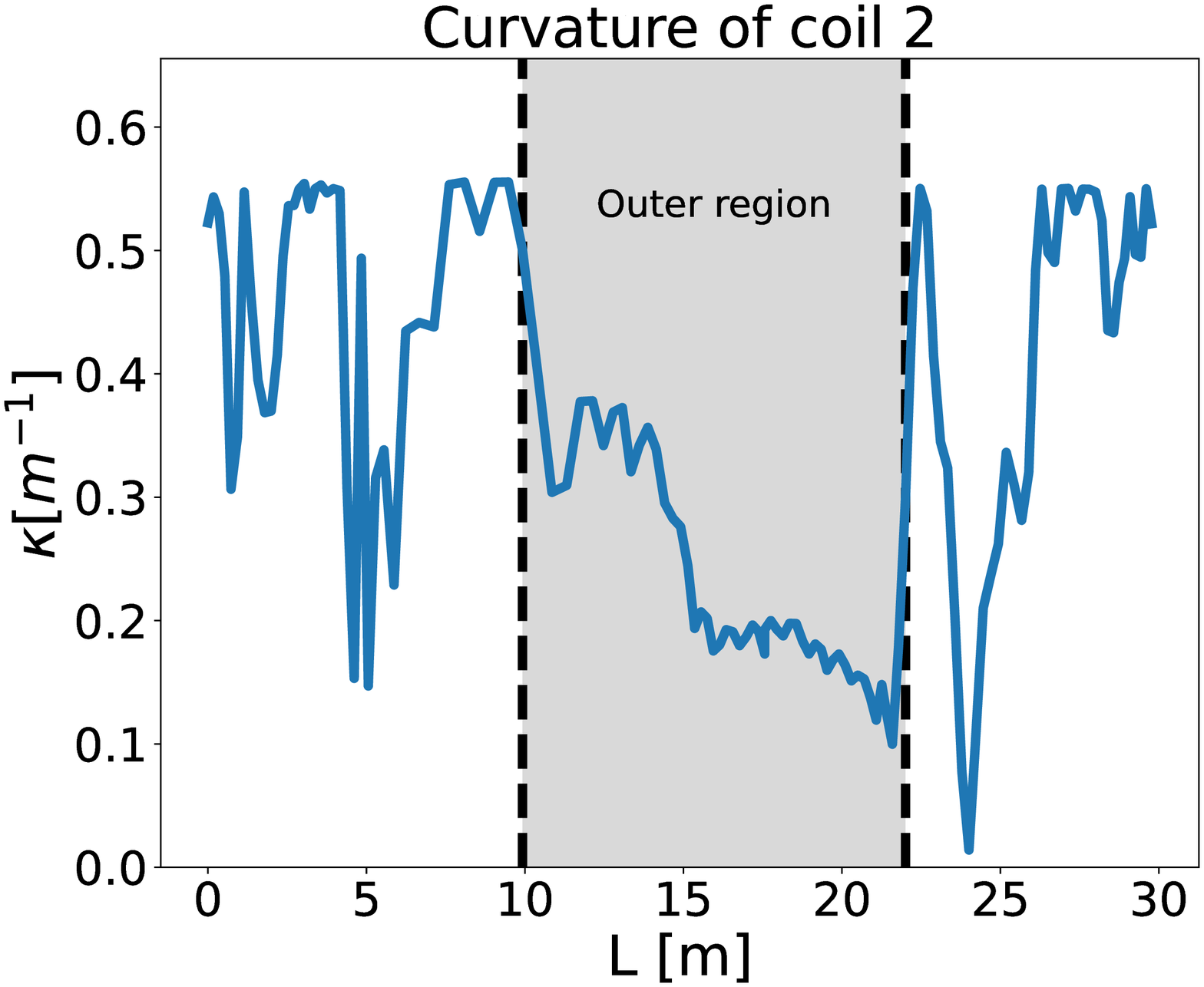}
\includegraphics[width=0.32\textwidth]{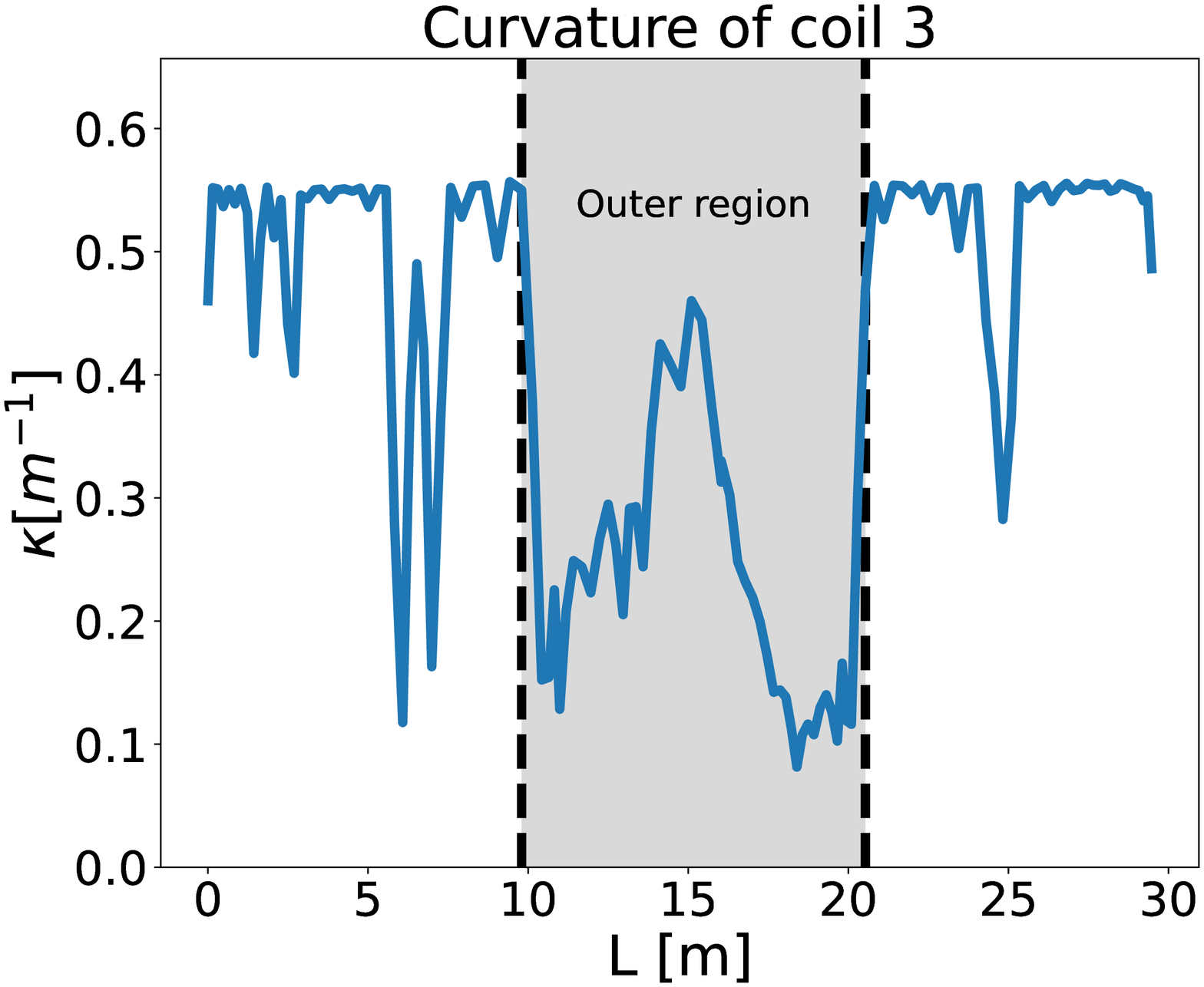}
\caption{Curvature $\kappa$ for each coil in the QAS prototype reactor. The gray areas indicate the regions of the optimized coils affected by the straight-out cost function. Coil 1 represents the left-most coil in Figure \ref{fig10} and coil 3 the right-most coil.}
\label{fig12}
\end{figure}

In figure \ref{fig11} the free-boundary VMEC results are shown and the flux surfaces follow the target surface closely along the entire field period, here represented by the section at the start of a period ($\phi = 0$) and at its middle point ($\phi = {\pi}/{3}$).
This suggests the coil set can produce the desired magnetic field supporting the target MHD equilibrium. 

\section{Conclusions and future work}\label{sec:concl}
A suitable spline representation for the coils in FOCUS was found and implemented in the code. This new representation allows for a more intuitive understanding of the optimization parameters and to constrain the resulting coils directly in real space.  A new cost function was  implemented to include an engineering constraint for straighter outer sections. These straighter sections have the potential to simplify the designs of future reactors and increase accessibility  on the outboard side, leaving space for ports and direct access to the modules of the blanket, while still providing good confinement.

The functional derivative used in this paper provides an easy approach to computing analytic derivatives with multiple parameterizations.
The functional derivative term only involves the geometry and has a close connection to the so-called ``shape gradient'' \cite{Landreman2018}.
Using the high flexibility that can be obtained by varying the parameters, the optimization procedure can be tweaked to design coils that balance the physical and engineering constraints for the experiment under study. 

In the future, the use of advanced optimization techniques implemented for the Fourier representation \cite{Zhu_02} and the coil sensitivity analysis \cite{Zhu_2018} can be extended to the spline one.
The advantages of a local representation could be exploited even further to specify the allowed interval for each control point coordinate and so impose arbitrary constraints on the shape of coils according to the user's needs, for example setting specific parts to be {\it exactly} straight and optimizing only the remaining part of the coils.

\ack
The authors thank T. Kruger for providing the HSX data, N. Pomphrey \& T. Brown for providing the QAS reactor data. This work was made possible by funding from the Department of Energy for the Summer Undergraduate Laboratory Internship (SULI) program. This work is supported by the U.S. DOE Contract No. DE-AC02-09CH11466.

\appendix
\section{Quasi-helical symmetry in HSX with the optimized coils}\label{sec:HSX_field_QH}
\begin{figure*}
\centering
\includegraphics[width=0.45\textwidth]{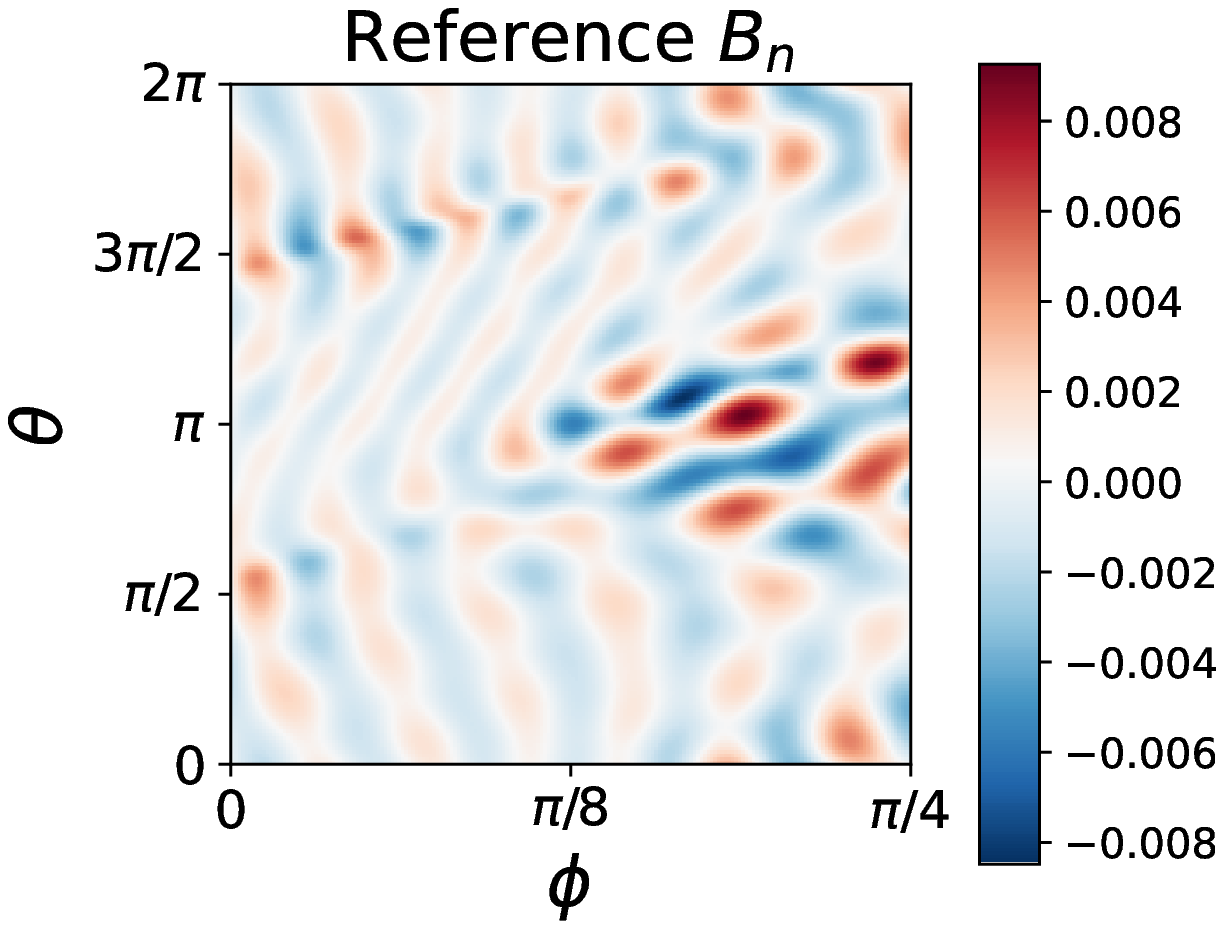}
\includegraphics[width=0.45\textwidth]{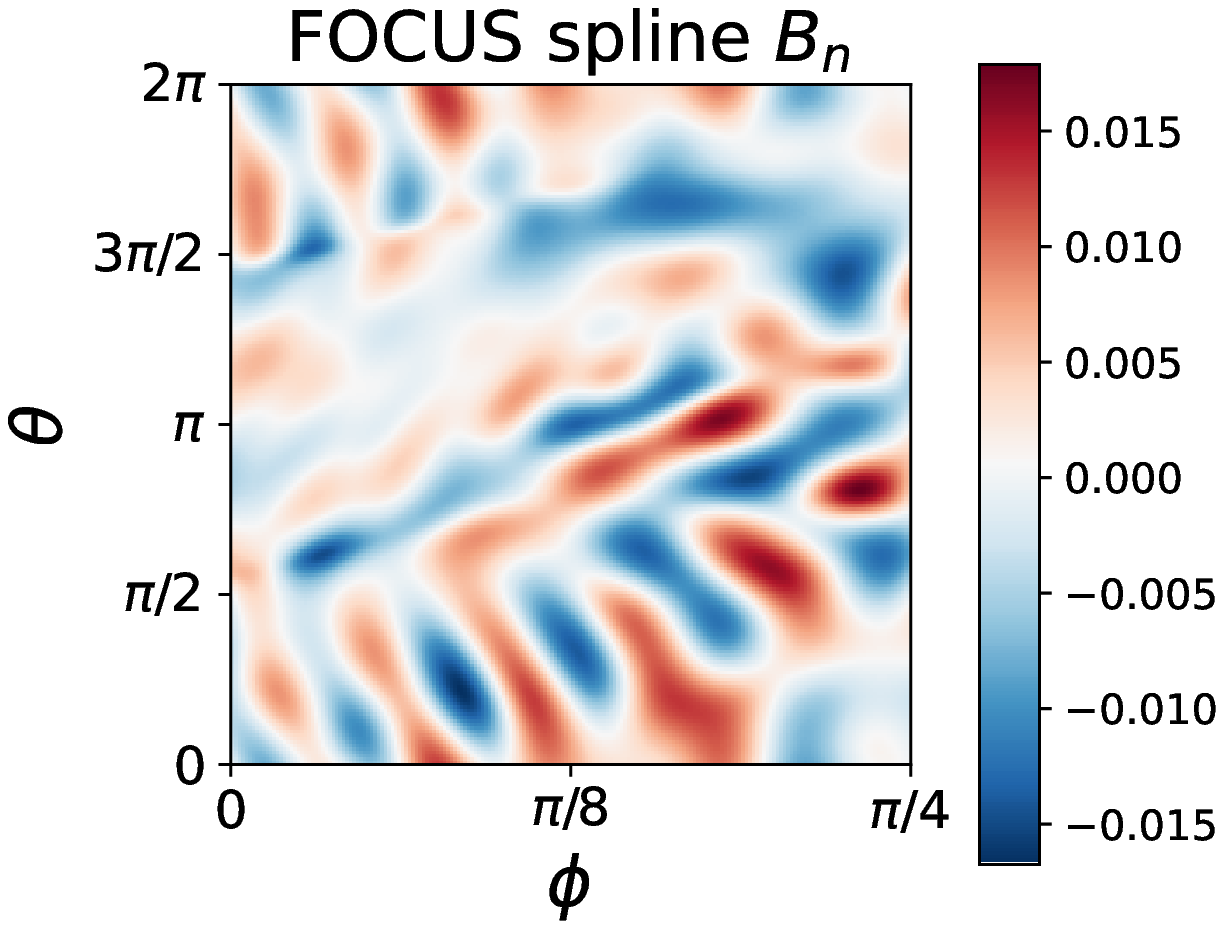} 
\caption{Normal magnetic field produced by the reference coil set (left) and the optimized coils (right) in HSX. $\theta$ and $\phi$ are the angles parameterizing the surface.}
\label{fig3}
\end{figure*}

\begin{figure}
\centering
\includegraphics[width=0.5\textwidth]{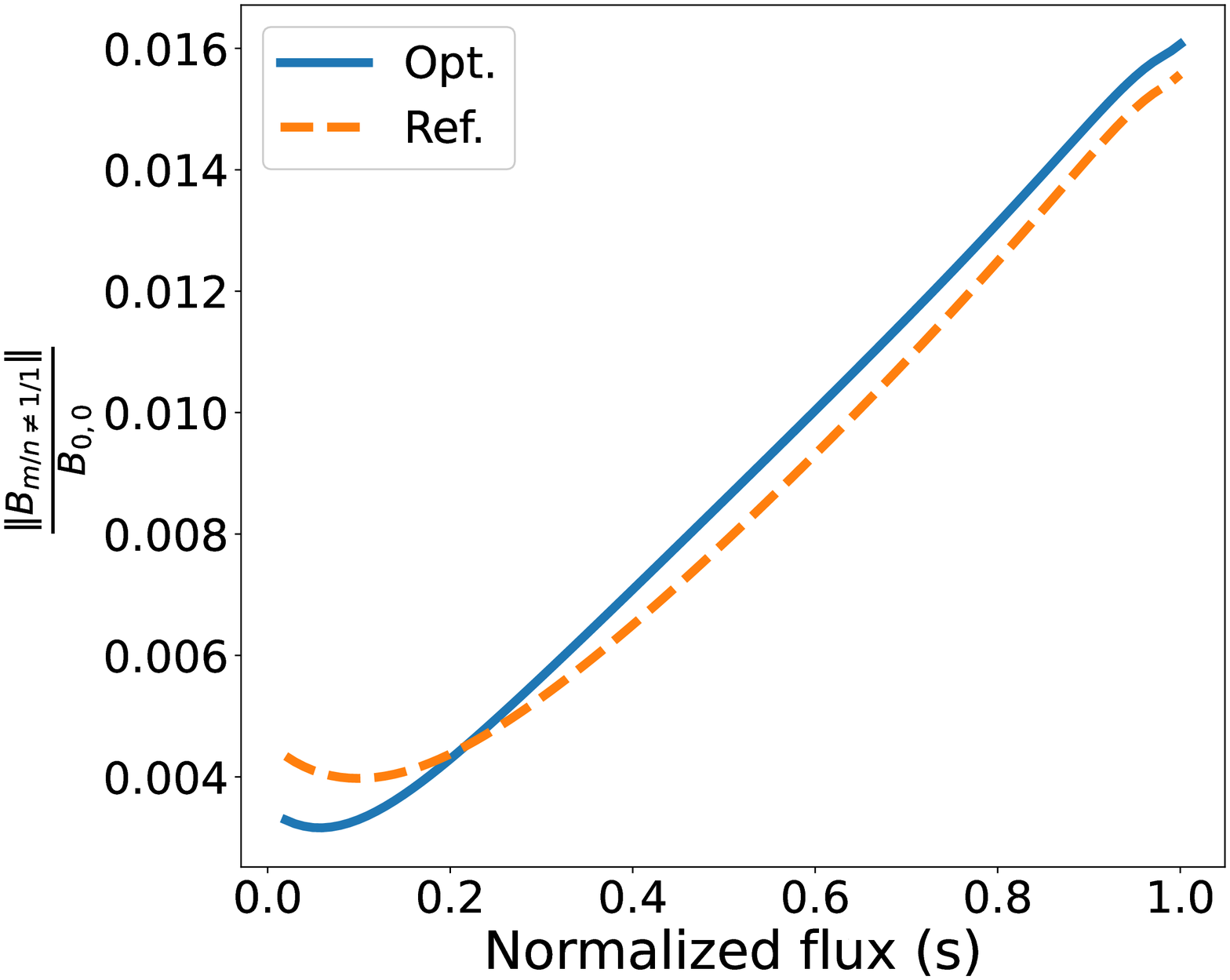} 
\caption{Non-helically symmetric fourier modes of the magnetic field in HSX}
\label{fig6}
\end{figure}

To verify the quality of the magnetic field produced by the new coil set, the normal magnetic field to the plasma surface is shown for the new configuration and the reference one in figure \ref{fig3}. In the new configuration a more visible coil ripple is present along with a slight increase of the normal field error. However, as shown in figure \ref{fig4}, the new coil set is still able to reproduce flux surfaces that follow the target surface.
Here, we further check the quality of quasi-helical symmetry, which is one of the main design goals. 
We compare the magnetic field components in the Boozer coordinates from the free-boundary VMEC equilibria.
As shown in figure \ref{fig6}, where the Fourier modes of the magnetic field different from the helical (1,1) mode are summed up, a good quasi-helical symmetry is maintained in both equilibria and the difference between the two configurations is negligible compared to the total value of the field.

\section*{References}
\bibliographystyle{iopart-num}
\bibliography{bib}

\end{document}